\documentclass[twocolumn,twoside,slac_two]{revtex4}
\usepackage{graphicx}
\usepackage{fancyhdr}
\def\DZero{DZero}
\def\calB{${\cal B}($}

\def\BsDsK{$B^0_s\to D_s^+ \, K^-$}
\def\BsDspi{$B^0_s\to D_s^- \, \pi^+$}
\def\BsDs1{$B_s^0\to D_{s1}^-(2536)\, \mu^+ \, \nu_\mu \, X$}
\def\BsDsDs{$B_s^0 \to D_s^- \, D_s^+$}
\def\BsDssDss{$B_s^0 \to D_s^{(*)} \, D_s^{(*)}$}
\def\BsDsspDssm{$B_s^0 \to D_s^{*+} \, D_s^{*-}$}
\def\BsDsspDsm{$B_s^0 \to D_s^{*+} \, D_s^{-}$}
\def\BsDspDsm{$B_s^0 \to D_s^{+} \, D_s^{-}$}
\def\Bsphig{$B_s^0\to\phi \, \gamma$}
\def\Bsgg{$B_s^0\to\gamma\, \gamma$}
\def\Lbpk{$\Lambda_b^0 \to p \, K^-$}
\def\Lbppi{$\Lambda_b^0 \to p \, \pi^-$}

\begin{document}



\title{Decays of $B_s^0$ Mesons and $b$ Baryons:\\
A Review of Recent First Observations and Branching Fractions}

%

\author{Andreas Warburton \\ (on behalf of the Belle, CDF, and
  \DZero\ Collaborations)\vspace{0.1in}}
\affiliation{McGill University, Montr\'{e}al, Qu\'{e}bec, Canada}

\begin{abstract}
Recent rate measurements of $B_s^0$ mesons and $\Lambda_b^0$ baryons
produced in $\sqrt{s} = 1.96~$TeV proton-antiproton and $\Upsilon(5S)$
electron-positron collisions are reviewed, including the first
observations of six new decay modes: \BsDsK\ (CDF), \BsDsDs\ (CDF),
\BsDs1\ (\DZero), \Bsphig\ (Belle), \Lbppi\ (CDF), and \Lbpk\ (CDF).
Also examined are branching-fraction measurements or limits for the
\BsDssDss\ modes (Belle, CDF, and \DZero), the \Bsgg\ radiative
penguin decay (Belle), and three two-body charmless $B_s^0$ meson
decay channels (CDF).  Implications for the phenomenology of
electroweak and QCD physics, as well as searches for physics beyond
the Standard Model, are identified where applicable.
\end{abstract}

\maketitle

\thispagestyle{fancy}


\section{Introduction}

Heavier $b$-flavoured hadrons represent a fecund source of particle
physics.  While the rich interplay between electroweak and
non-perturbative strong effects typically poses formidable
experimental and theoretical challenges, decays of hadrons with masses
at the frontiers of Standard Model spectroscopy constitute an exciting
proving ground for effective theories, QCD factorization and lattice
methods, as well as potential models.  Moreover, such heavy hadronic
states present opportunities to uncover real or constrain hypothetical
new physics lying beyond the Standard Model.

The measurement of observables from $b$ baryons and strange or charmed
$B$ mesons is complementary to the wealth of physics that the {\sc
  BaBar}, Belle, and CLEO collaborations have harvested from
$e^+e^-$ colliders operating at the $\Upsilon(4S)$ open-beauty
threshold.  Comparisons of heavy $b$-hadron decays to the analogous
non-strange $B_{u,d}$ ($B^{+,0}$) decays can yield advantages that
include cancellations of hadronic uncertainties, tests of $SU(3)$
flavour symmetry, decay-amplitude disentanglement, and improved access
to fundamental electroweak parameters of Nature.

This paper reviews recent rate measurements of $B_s^0$ mesons and
$\Lambda_b^0$ baryons produced in $\sqrt{s} = 1.96~$TeV
proton-antiproton and $\Upsilon(5S)$ electron-positron collisions at
the Fermilab Tevatron (USA) and KEKB (Japan) accelerator facilities,
respectively.  Described are first observations of \BsDsK\ and
\BsDsDs\ decays\footnote{Charge conjugate decays are implied
  throughout.} by CDF; recent results on \BsDssDss\ decays and
worldwide status from Belle, CDF, and \DZero; the first observation of
\BsDs1\ by \DZero; the first observation of the \Bsphig\ mode and a
search for \Bsgg\ decays by Belle; three recently updated measurements
of charmless two-body $B_s^0$ meson decays by CDF; and first
observations of \Lbpk\ and \Lbppi\ decays by CDF.

\section{First \BsDsK\ Observation}
\label{BsDsK}
\begin{figure}[h]
\centering
\includegraphics[width=80mm]{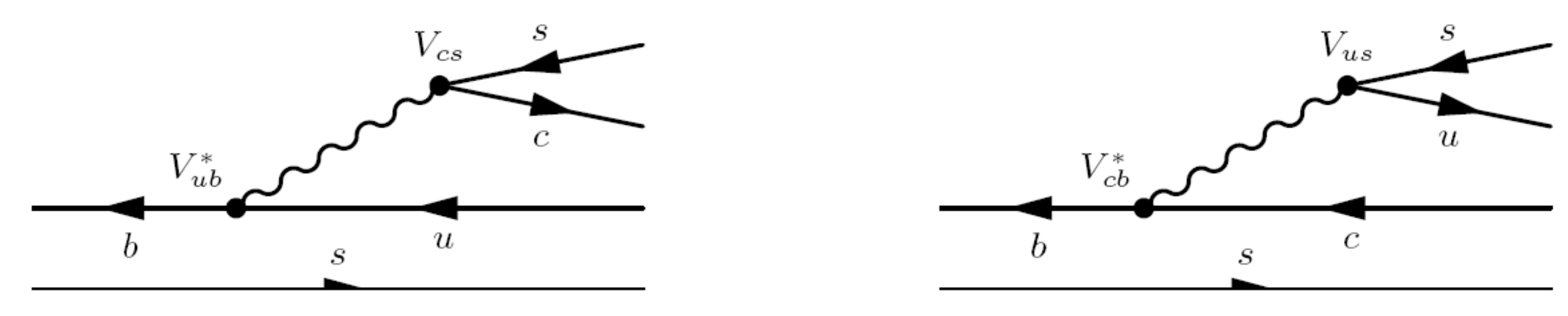}
\caption{Upper- and lower-vertex charm diagrams contributing to \BsDsK
  decay~\cite{CDF:BsDsK}.} \label{fig:BsDsK_Diagrams}
\end{figure}

The Cabibbo-suppressed \BsDsK\ mode, as indicated in
Fig.~\ref{fig:BsDsK_Diagrams}, has contributions from both upper- and
lower-vertex charm such that the same $B^0_s$ parent state can decay
both to $D_s^+\,K^-$ and $D_s^-\,K^+$ final states, as distinct from
the analogous $B^0\to D^- \, K^+$ mode.  Due to the fact that these
decay amplitudes can interfere through $B_s$ mixing, the branching
fraction relative to that of the Cabibbo-favoured reference mode,
${\cal B}($\BsDsK$)/{\cal B}($\BsDspi$)$, can deviate significantly
higher or lower than the analogous ratio of non-strange branching
fractions, ${\cal B}(B^0 \to D^-\, K^+)/{\cal B}(B^0 \to D^-\,
\pi^+)$.  As is also evident from Fig.~\ref{fig:BsDsK_Diagrams}, the
relative weak phase between the two final charge states is the angle
$\gamma$ ($\phi_3$) of the CKM unitarity triangle, meaning that a
flavour-tagged time-dependent analysis of \BsDsK\ could yield a
theoretically clean~\cite{Aleksan} measurement of $\gamma$ ($\phi_3$).
The possibility of an untagged approach relying on a significant
lifetime difference between the $B_s$ $CP$ eigenstates has also been
posited~\cite{Dunietz}.

The CDF collaboration recently reported~\cite{CDF:BsDsK} the first
observation of \BsDsK\ decays and the first measurement of the ${\cal
  B}($\BsDsK$)/{\cal B}($\BsDspi$)$ branching-fraction ratio using
$D_s^-\to\phi\,\pi^-$ decays in 1.2~fb$^{-1}$ of $p\bar{p}$ collisions
at $\sqrt{s} = 1.96~$TeV.  The analysis made profitable use of CDF's
Level~2 Silicon Vertex 2-Track Trigger~\cite{CDF:TTT}, which could
identify vertices from track pairs with significant impact parameters
(between 0.12 and 1~mm), and therefore functioned as a valuable online
means to identify relatively pure samples of hadronically decaying
heavy-flavour hadrons.  Two separate control samples were studied
simultaneously to form analogous $D^{*-}\, K^+ / D^{*-}\, \pi^+$ and
$D^{-}\, K^+ / D^{-}\, \pi^+$ ratios as cross checks.  The selection
criteria~\cite{CDF:BsDsK} involved no explicit particle identification
requirements such that tracks were taken to be either kaons or pions to
match the reconstruction hypothesis.

The principal challenge in this analysis was the disentanglement,
using a multivariate maximum-likelihood fit, of the numerous
components entering the $D_s^-\,\pi^+$ data sample.  A simultaneous
fit was performed in terms of the $D_s^-\,\pi^+$ invariant mass and a
particle-identification variable $Z$ describing the non-$D_s^-$
daughter track, where $Z\equiv \log\left[dE/dx({\rm measured}) /
  dE/dx({\rm expected\ for\ }\pi) \right]$ such that it is normally
distributed.  Mass templates were created from sizeable Monte Carlo
samples to be used as probability density functions (PDFs) in the fit.
Combinatorial background due to real $D$ mesons was estimated using
wrong-sign data, and the $dE/dx$ $Z$-variable templates were derived
largely from inclusive $D^*$ data.

\begin{figure}[h]
\centering
\includegraphics[width=80mm]{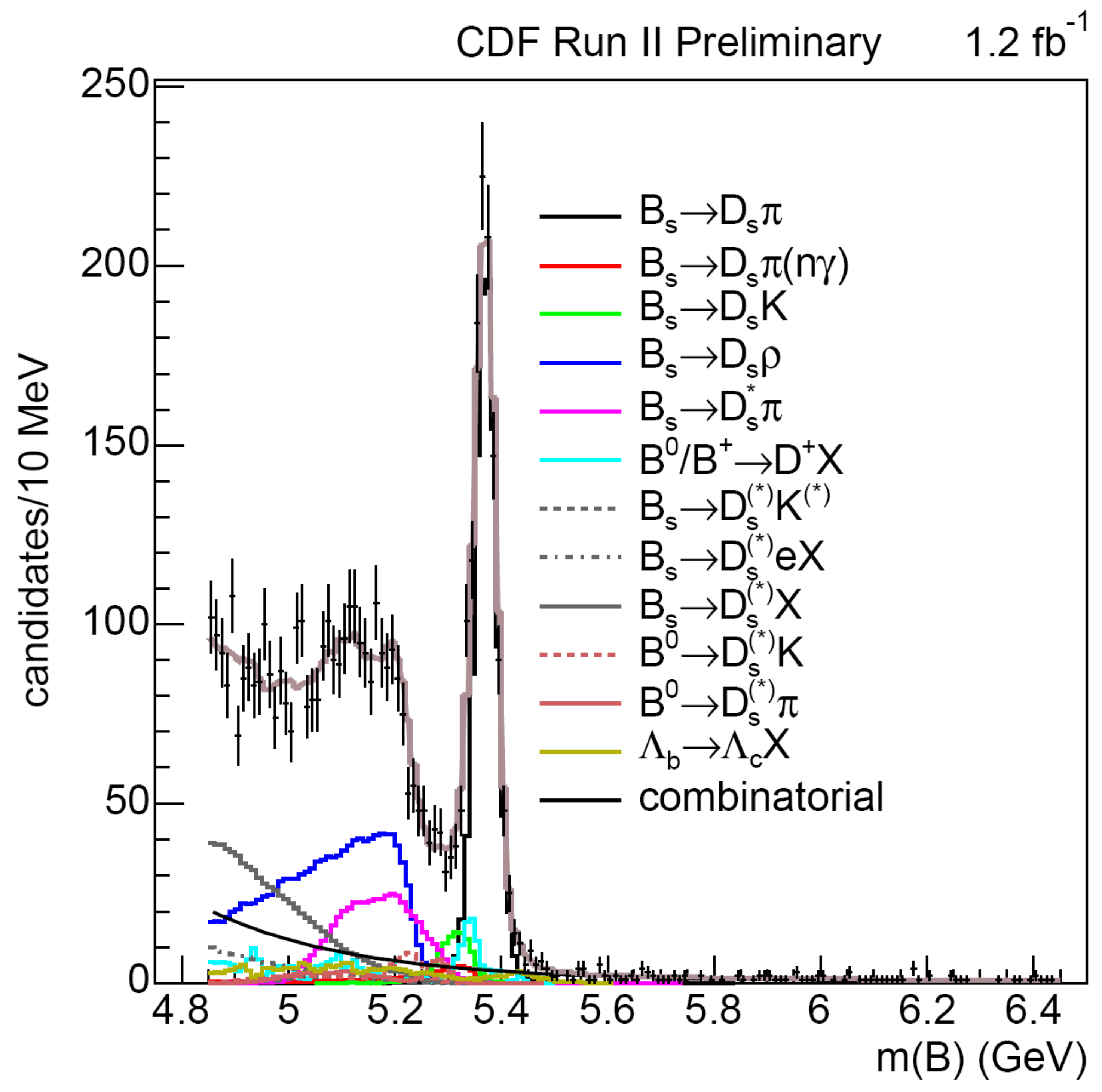}
\caption{Invariant mass projections of the likelihood fit results for
  \BsDspi\ candidates~\cite{CDF:BsDsK}.  The fit result for the signal
  \BsDsK\ distribution is indicated by the green
  histogram.} \label{fig:BsDsK_mass_projection}
\end{figure}

\begin{figure}[h]
\centering
\includegraphics[width=80mm]{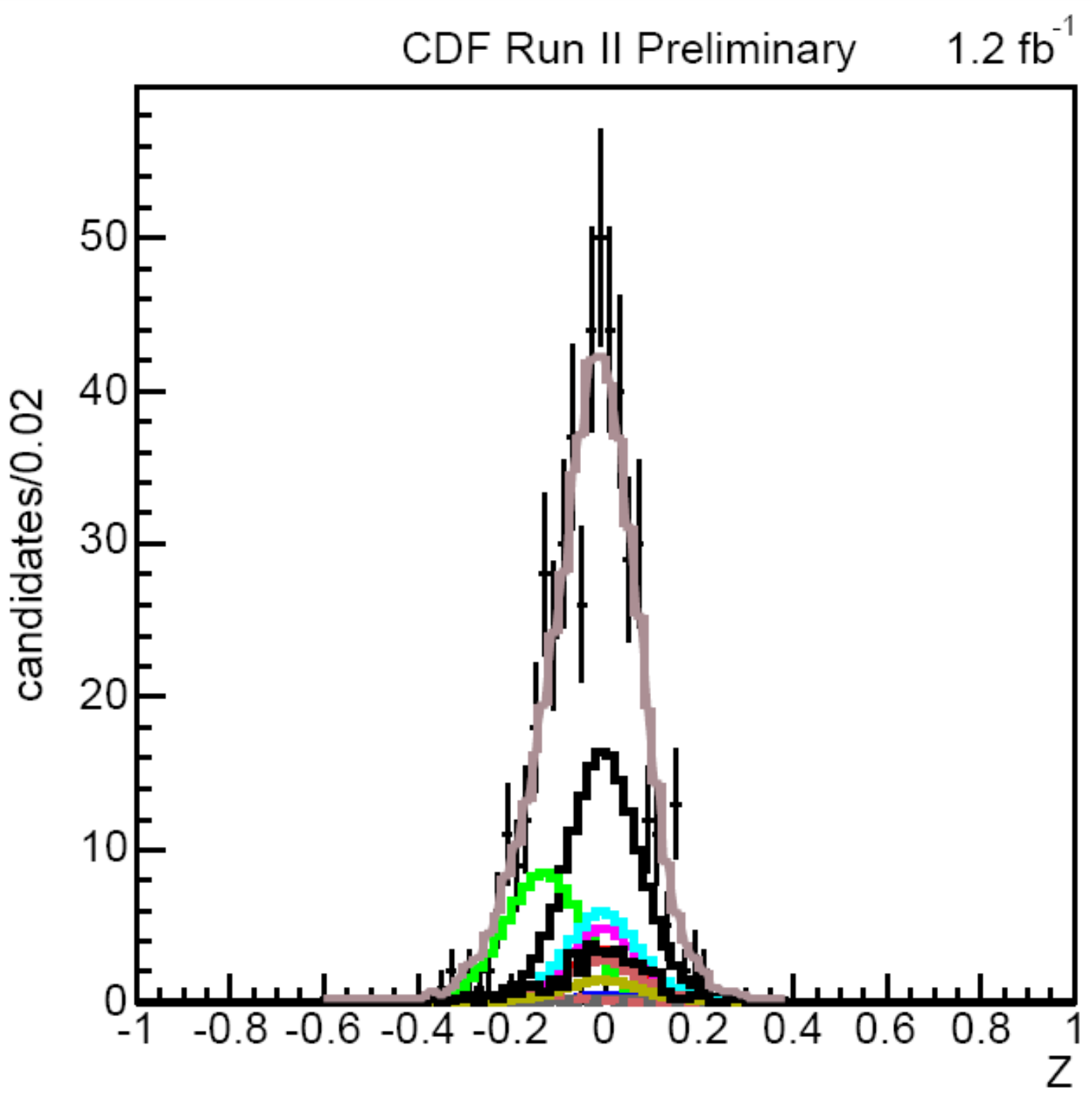}
\caption{$Z$ particle-identification projections of the likelihood fit
  results for \BsDspi\ candidates for the $D_s\,\pi$ mass window
  corresponding to the $D_s\, K$ signal region ($5.26 -
  5.35$~GeV/$c^2$)~\cite{CDF:BsDsK}.  The colour coding is identical
  to that shown in Fig.~\ref{fig:BsDsK_mass_projection}, and the
  signal \BsDsK\ distribution is again shown in
  green.} \label{fig:BsDsK_Z_projection}
\end{figure}

The fit, for which care was taken to treat the $D\,\pi$ radiative tail
as a free parameter and correlations amongst the fit parameters were
taken into account, identified $109\pm 19$ $D_s\, K$ signal
candidates with a statistical significance of 7.9 standard deviations.
Invariant mass and $Z$ projections from the likelihood fit are
indicated in Figs.~\ref{fig:BsDsK_mass_projection}
and~\ref{fig:BsDsK_Z_projection}, respectively.

The resulting ratio of branching fractions was measured to be

\begin{equation}
\frac{{\cal B}(B^0_s\to D_s^+ \, K^-)}{{\cal B}(B^0_s\to D_s^- \, \pi^+)} =
0.107 \pm 0.019 ({\rm stat.}) \pm 0.008 ({\rm syst.}),
\label{eqn:BsDsK_BsDspi}
\end{equation}

for which the largest source of systematic uncertainty arose from the
treatment of the $Z$ variable templates.  The result expressed in
Eqn.~\ref{eqn:BsDsK_BsDspi} is statistically compatible with the
analogous $B^0$ branching-fraction ratio, indicating that interference
effects are not yet observable.  A recent CDF
measurement~\cite{CDF:BsDspi} also exists for the branching fraction of the
normalization mode, ${\cal B}($\BsDspi$)$, and may be used to extract
an absolute branching fraction for \BsDsK\ decays.

\section{Recent \BsDssDss\ Decay Results}
\label{BsDssDss}
The suite of $B_s^0$ decays denoted by \BsDssDss, representing the
three decay modes \BsDsspDssm, \BsDsspDsm, and \BsDspDsm, is
considered to be the principal contributor to the rate difference
$\Delta\Gamma_s^{CP}$ between the odd and even $B_s^0$ $CP$
eigenstates.  Mixing in the $B_s^0$ system has recently been observed,
thus providing a precise measurement of the mass difference of the
$B_s^0$ mass eigenstates~\cite{CDF:BsMixing}.  To a good
approximation, the mass eigenstates are expected to be eigenstates of
$CP$ such that their rate difference is equivalent to
$\Delta\Gamma_s^{CP}$.  Assuming \BsDssDss\ decays dominate the $b\to
c\bar{c}s$ quark-level transitions, which are purely $CP$-even, the
branching fraction is related to $\Delta\Gamma_s^{CP}$ by ${\cal
  B}($\BsDssDss$)=\Delta\Gamma_s^{CP}/2\Gamma_s$ in the Standard
Model~\cite{BsDssDss_Theory}, which regards the $CP$-violating mixing
phase as negligible.  Studies of ${\cal B}($\BsDssDss$)$, modulo
theoretical assumptions about dominant $CP$ content, can therefore
reveal departures from the Standard Model in the form of new physics
that renders the $CP$-violating mixing phase non-zero.

Although the overall \BsDssDss\ rates are expected to be large,
studies of these modes are experimentally challenging due to the small
fully hadronic charm-meson branching fractions.  Two recent studies at
the Tevatron are reviewed here, followed by a brief summary of the
worldwide status of \BsDssDss\ decay measurements.

\subsection{\DZero\ Inclusive ${\cal B}($\BsDssDss$)$}
\label{DZero:BsDssDss}
The \DZero\ collaboration recently presented a preliminary inclusive
measurement of \calB\BsDssDss$)$ based on a 2.8~fb$^{-1}$ sample of
$p\bar{p}$ collisions at $\sqrt{s} = 1.96~$TeV~\cite{DZero:BsDssDss}.
No distinction was made between the orbitally excited and ground-state
$D_s^{(*)}$ mesons; the analysis therefore sought to identify
correlations between final-state particles in a pair of candidate
$D_s^{(*)}$ decays, the first involving a $\phi_1\,\pi$ final state
and the second involving a $\phi_2\,\mu\,\nu$ final state, where
$\phi_1$ and $\phi_2$ denote two mutually exclusive $\phi\to K^+\,
K^-$ candidate reconstructions in the same event.  The majority of
candidate events was collected using single-muon triggers by selecting
a common sample~\cite{DZero:BsDssDss} containing muon track candidates
and $D_s\to\phi_1\,\pi$ candidates.

\begin{figure}[h]
\centering
\includegraphics[width=80mm]{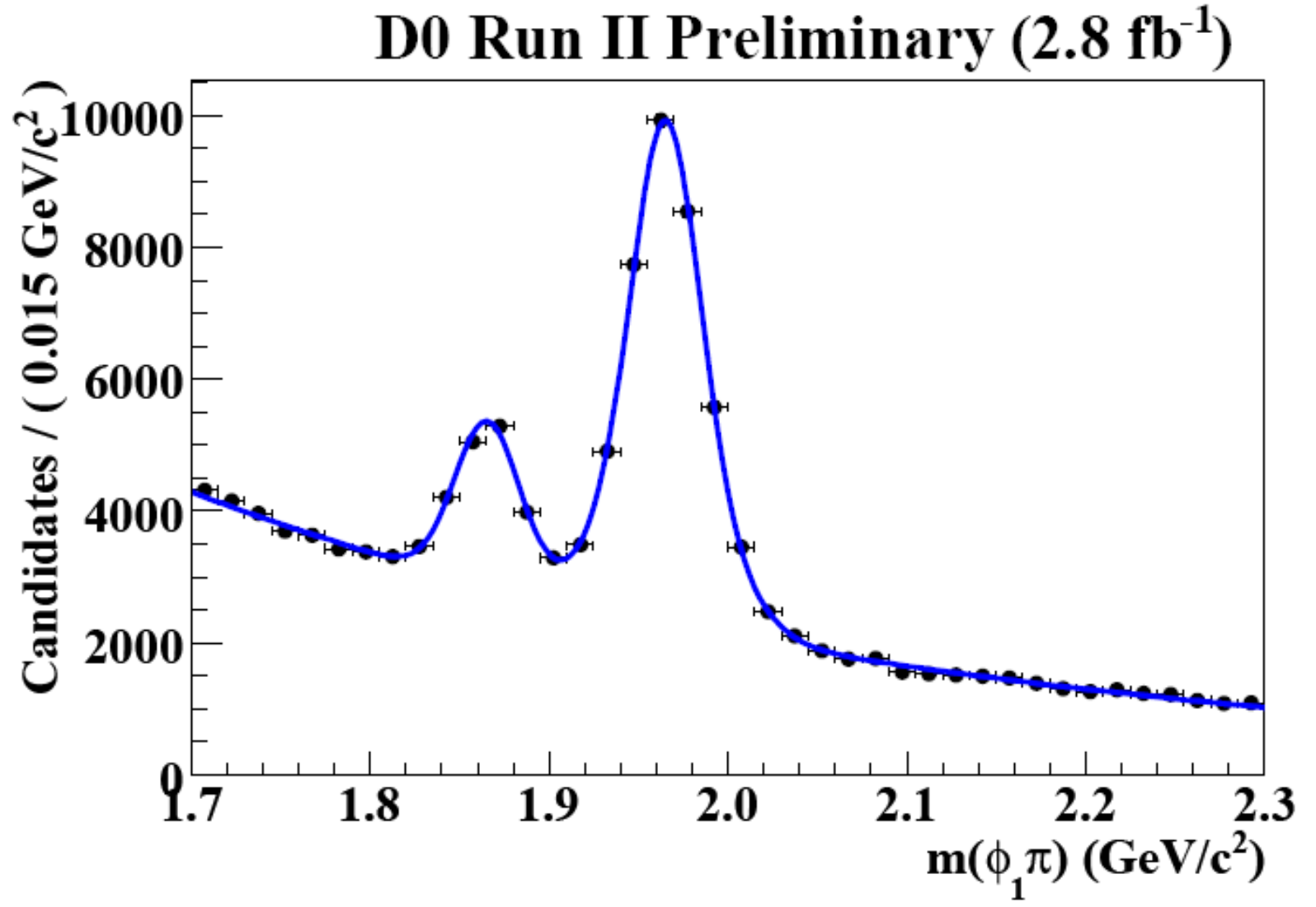}
\caption{Inclusive $D^{(*)}$ candidate reconstruction in the $\mu$
  sample, represented by the $\phi_1\,\pi$ invariant mass distribution
  corresponding to the $K^+\, K^-$ signal region, $1.01 < m(K^+\, K^-)
  < 1.03$~GeV/$c^2$~\cite{DZero:BsDssDss}.  The two peaks represent
  the observed $D^\pm$ and $D_s^\pm$ inclusive candidate
  yields.} \label{fig:DZeroPhi1pi}
\end{figure}

\begin{figure}[h]
\centering
\includegraphics[width=80mm]{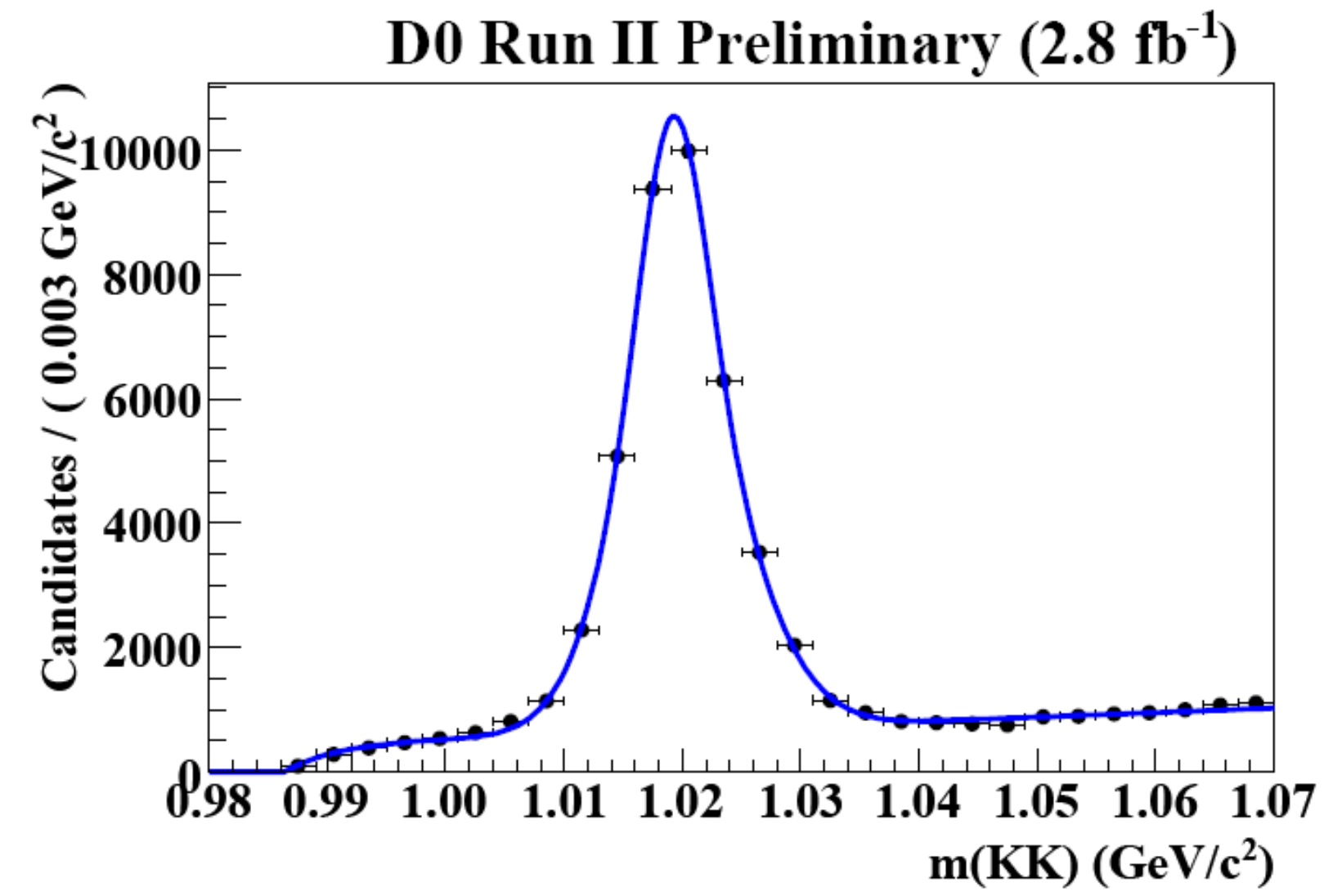}
\caption{Inclusive $K^+\, K^-$ invariant mass distribution for
  $\phi_1$ candidates corresponding to the $\phi_1\,\pi$ signal
  region, $1.92 < m(\phi_1\,\pi) <
  2.00$~GeV/$c^2$~\cite{DZero:BsDssDss}.} \label{fig:DZeroPhi1}
\end{figure}

A binned likelihood fit identified the inclusive number of $D_s\,\mu$
candidates as $28\,680\pm 288 ({\rm stat.})$, taking into account
possible $D^\pm$ contributions and combinatorial backgrounds.
Figs.~\ref{fig:DZeroPhi1pi} and~\ref{fig:DZeroPhi1} illustrate the
invariant mass distributions of the initial $\phi_1\,\pi$ and $\phi_1$
candidate reconstructions, respectively.

\begin{figure}[h]
\centering
\includegraphics[width=80mm]{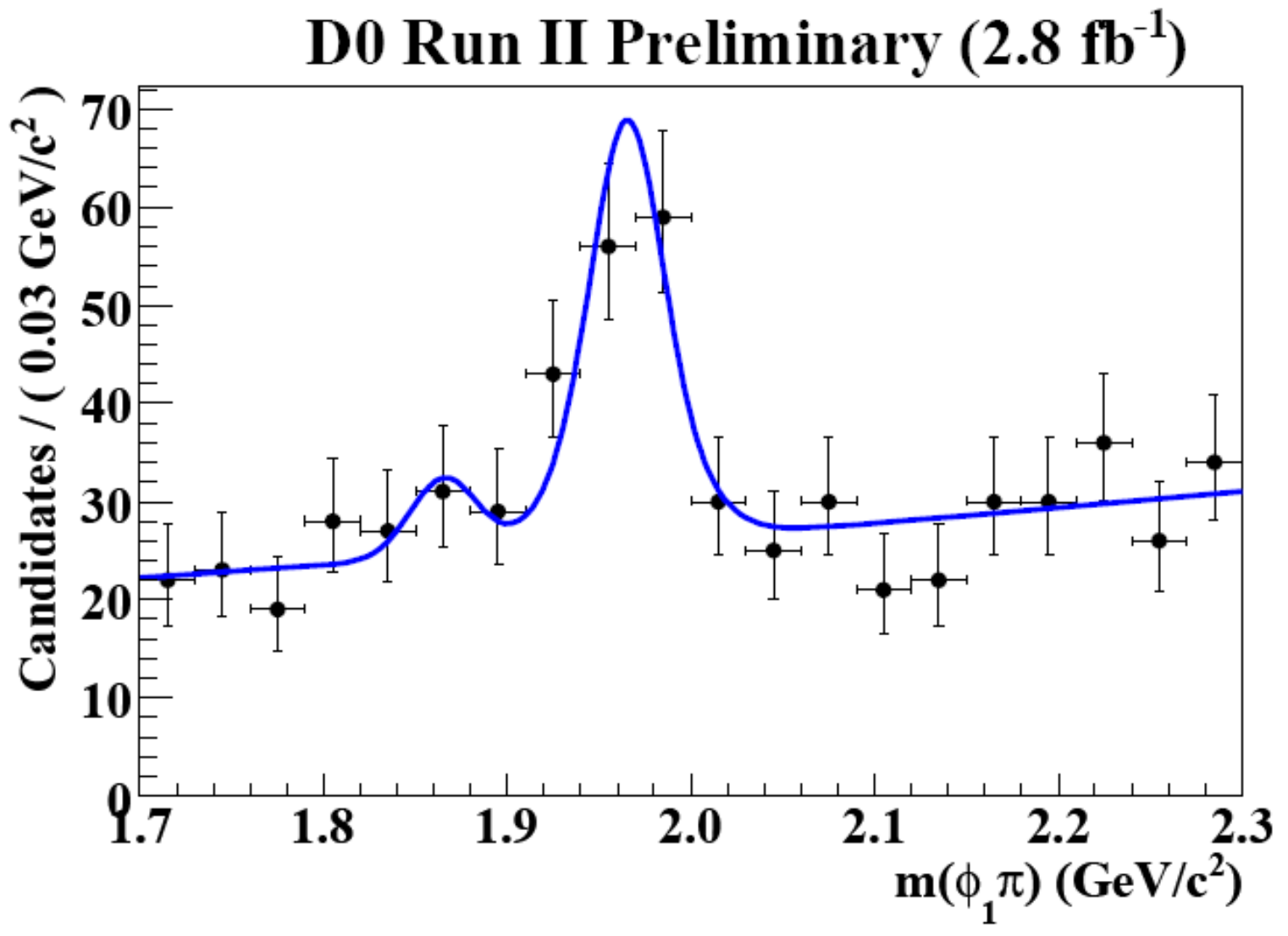}
\caption{Inclusive $D^{(*)}$ candidate reconstruction in the $\mu$
  sample with the additional $\phi_2\,\mu$ requirement applied,
  represented by the $\phi_1\,\pi$ invariant mass distribution
  corresponding to the $K^+\, K^-$ signal region, $1.01 < m(K^+\, K^-)
  < 1.03$~GeV/$c^2$~\cite{DZero:BsDssDss}.  The two peaks represent
  the observed $D^\pm$ and $D_s^\pm$ inclusive candidate
  yields.} \label{fig:DZeroPhi2pi}
\end{figure}

\begin{figure}[h]
\centering
\includegraphics[width=80mm]{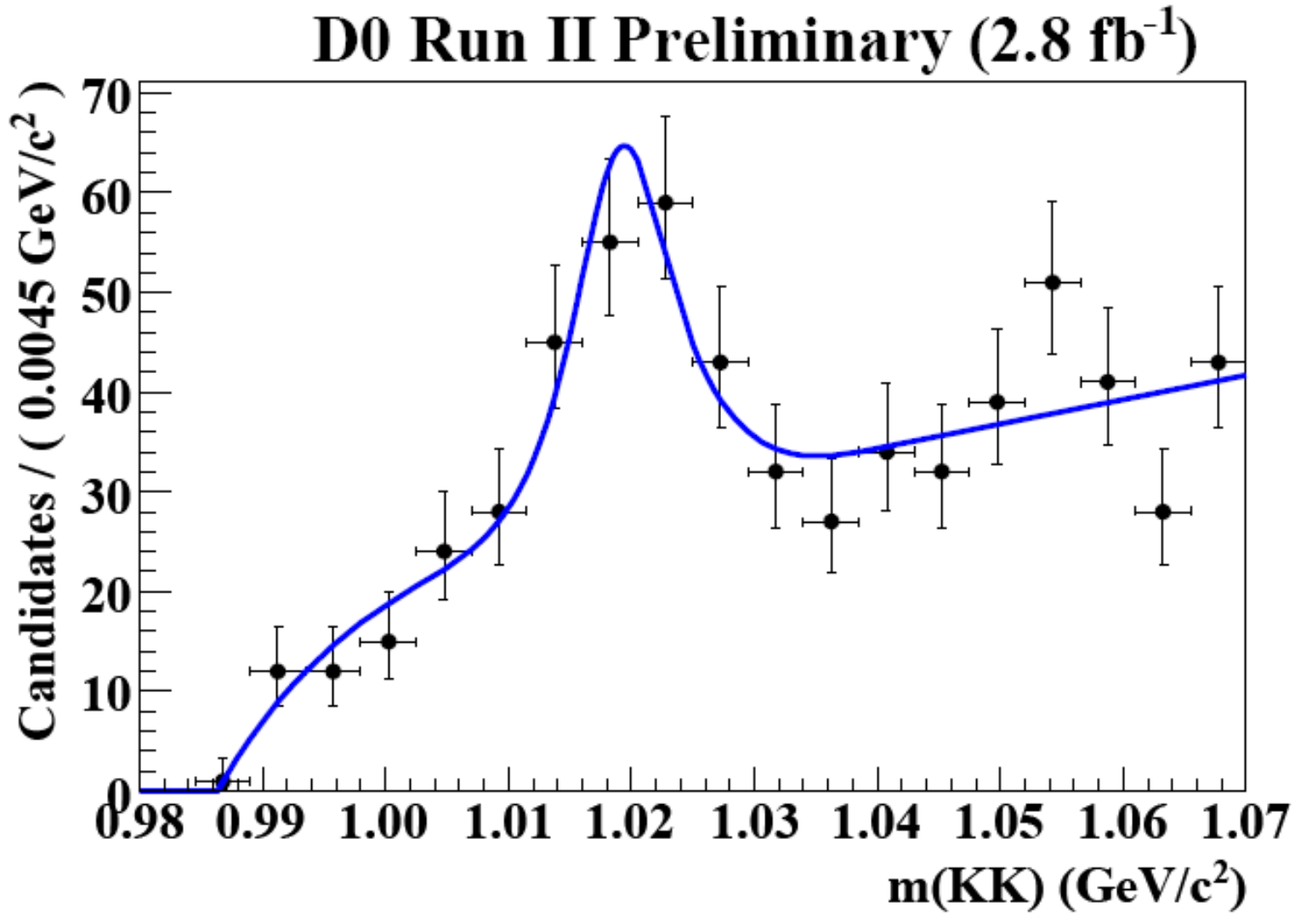}
\caption{Inclusive $K^+\, K^-$ invariant mass distribution for
  $\phi_2$ candidates composed of tracks other than those used to form
  $\phi_1$ candidates~\cite{DZero:BsDssDss}.} \label{fig:DZeroPhi2}
\end{figure}

The presence of a second distinct $D_s^{(*)}$ candidate was determined
by searching for a different $K^+\, K^-$ candidate system representing
the $\phi_2$ component, similarly selected but required to share a
vertex with the muon candidate, which is taken by definition as a
daughter of the second $D_s^{(*)}$ candidate.  A two-dimensional
unbinned likelihood fit was used to identify the correlation between
the $\phi_1\,\pi$ and $\phi_2$ candidates, yielding $31.0\pm 9.4 ({\rm
  stat.})$ $\phi_2\,\mu$ candidates with a statistical significance of
3.7 standard deviations.  Figs.~\ref{fig:DZeroPhi2pi}
and~\ref{fig:DZeroPhi2} illustrate the result of modifying
Fig.~\ref{fig:DZeroPhi1pi} with the $\phi_2$ candidate requirement and
the inclusive $\phi_2$ invariant mass distribution, respectively.  The
resultant branching-fraction measurement is extracted by normalizing
to the reference mode $B^0_s\to D_s^{(*)}\,\mu\,\nu$, the branching
fraction for which contributes the largest systematic uncertainty to
the result; the absolute \BsDssDss\ branching fraction is summarized
in Sect.~\ref{worldwide} below.

\subsection{CDF Exclusive ${\cal B}$(\BsDsDs$)$}
\label{CDF:BsDsDs}
The first observation of the exclusive decay \BsDsDs\ was made by the
CDF Collaboration in 355~pb$^{-1}$ of $p\bar{p}$ collisions at
$\sqrt{s} = 1.96~$TeV~\cite{CDF:BsDsDs}.  The \BsDsDs\ mode is deemed
to be purely $CP$ even due to quark-level effects, meaning that the
$B_s^0$ parent is essentially always the light mass eigenstate in this
case.  The experimental approach used was to reconstruct candidates
exclusively by combining three different $D_s^+$ meson decay modes to
the $\phi\,\pi^+$, $\bar{K}^{*0}\,K^+$, and $\pi^+\,\pi^+\,\pi^-$
final states for one of the $B_s^0$ candidate daughters while
reconstructing the second daughter solely using the $\phi\,\pi^-$
mode.  Analogous $B^0\to D^-\, D_s^+$ decay candidates, with $D^-\to
K^+\,\pi^-\,\pi^-$, were reconstructed in parallel for normalization
purposes, with an attendant cancellation of several systematic
uncertainties.

\begin{figure}[h]
\centering
\includegraphics[width=80mm]{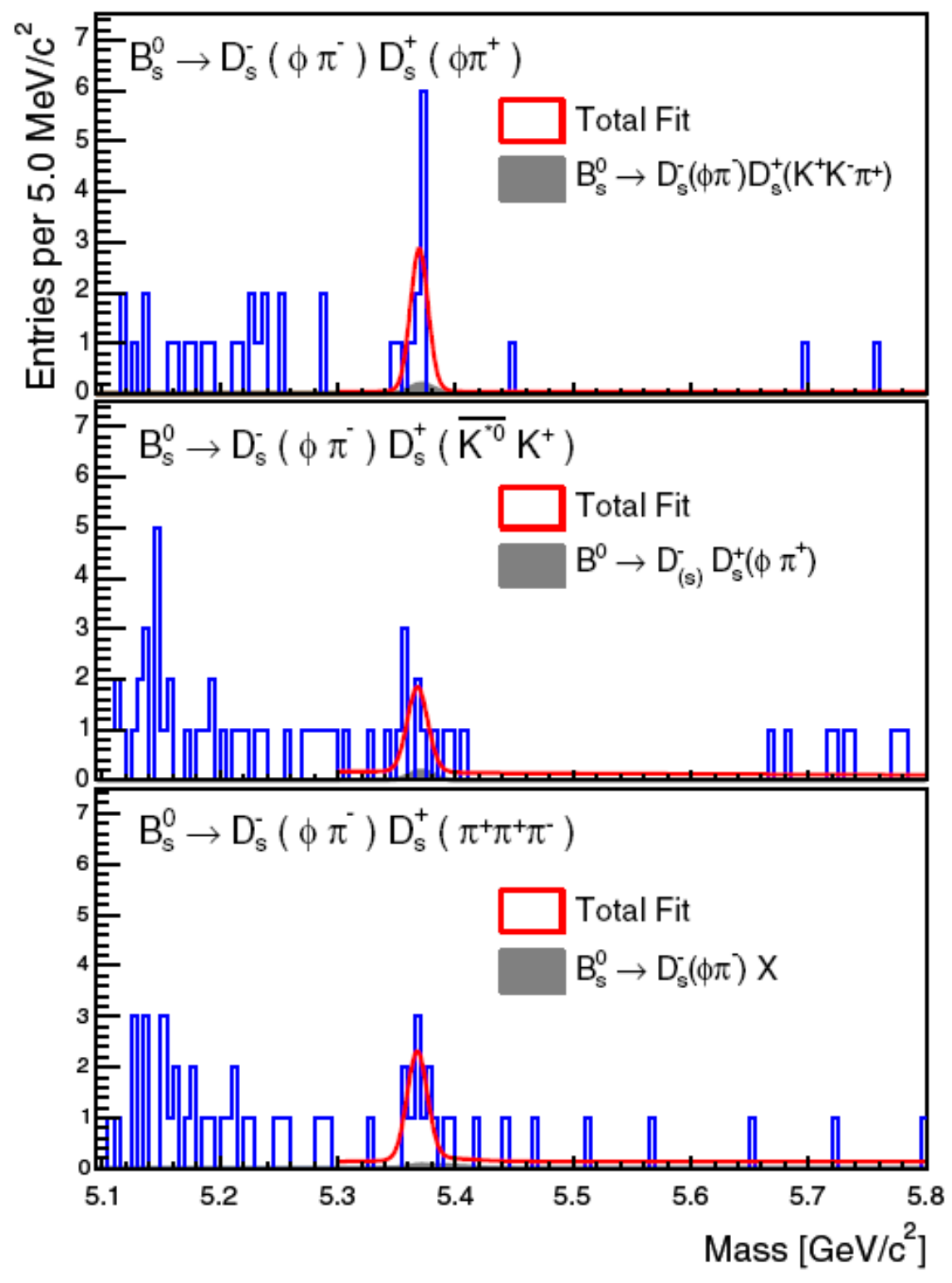}
\caption{Invariant mass distributions for \BsDsDs\ candidates
  (histogram), where $D_s^+\to\phi\,\pi^+$ (top),
  $D_s^+\to\bar{K}^{*0}\,K^+$ (middle), and $D_s^+\to
  \pi^+\,\pi^+\,\pi^-$ (bottom)~\cite{CDF:BsDsDs}.  The red curves
  describe the total fit result whereas the grey shaded regions
  represent the principal background component in each
  case.} \label{fig:CDF_BsDsDs}
\end{figure}

Strong signals with well understood backgrounds were reconstructed for
all three of the reference $B^0\to D^-\, D_s^+$ decay
modes~\cite{CDF:BsDsDs}.  Fig.~\ref{fig:CDF_BsDsDs} depicts the
invariant mass distributions and signal fits for the three
\BsDsDs\ decay channels.  The individual significances of the signal
reconstructed with the modes containing the $D_s^+\to\phi\,\pi^+$,
$D_s^+\to\bar{K}^{*0}\,K^+$, and $D_s^+\to \pi^+\,\pi^+\,\pi^-$ final
states were found to be 5.8, 3.4, and 4.4 standard deviations,
respectively.  The product of the three likelihoods was used to find a
combined result consistent with an observation of \BsDsDs\ decay with
7.5 standard deviations of significance.  After correcting the
relative event yields for relative acceptances and efficiencies, the
ratio of branching fractions was measured to be \calB\BsDsDs$) /
$\calB$B^0\to D^-\, D_s^+) = 1.44^{+0.48}_{-0.44}$, where
world-average measurements~\cite{PDG} were employed to correct for the
relative fragmentation fractions $f_d/f_s$ and daughter branching
fractions \calB$D^-\to
K^+\,\pi^-\,\pi^-)/$\calB$D_s^-\to\phi\,\pi^-)$, both of which
contribute prominently to the total systematic uncertainty of the
measurement.

\subsection{Absolute \BsDssDss\ Branching Fractions: Worldwide Status}
\label{worldwide}
The 3.7 standard deviation \DZero\ measurement~\cite{DZero:BsDssDss}
described in Sect.~\ref{DZero:BsDssDss} above yielded an absolute
branching fraction \calB\BsDssDss$) = 0.042\pm 0.015 ({\rm stat.}) \pm
0.017 ({\rm syst.})$, which confirms and is significantly more precise
than the PDG-updated~\cite{PDG} measurement of the first observation
by ALEPH~\cite{ALEPH:BsDssDss}, $0.12\pm 0.05 ({\rm stat.})
\ ^{+0.10}_{-0.04}({\rm syst.})$.

The 7.5 standard deviation CDF measurement~\cite{CDF:BsDsDs} described
in Sect.~\ref{CDF:BsDsDs} above yielded an absolute branching fraction
\calB\BsDsDs$) = \left( 9.4^{+4.4}_{-4.2}\right) \times 10^{-3}$,
representing the first observation of this mode.

The Belle collaboration has also recently placed limits on the three
exclusive decay constituents of the \BsDssDss\ system of modes.  The
results, which are based on a 1.86~fb$^{-1}$ data sample obtained at
the $\Upsilon(5S)$ resonance in $e^+e^-$ collisions at the KEKB
asymmetric collider, are \calB\BsDsDs$) < 6.7$\%, \calB\BsDsspDsm$) <
12.1$\%, and \calB\BsDsspDssm$) < 25.7$\%, each at 90\% confidence
level~\cite{Belle:BsDssDss}.

By invoking certain theoretical assumptions, measurements of
\calB\BsDssDss$)$ and \calB\BsDsDs$)$ can readily be applied to
estimating or placing a lower bound on $\Delta\Gamma_s^{CP}/\Gamma_s$,
as discussed at the beginning of Sect.~\ref{BsDssDss}.  Both the
\DZero\ and CDF branching-fraction measurements suggest values of
$\Delta\Gamma_s^{CP}/\Gamma_s$ consistent with the Standard Model,
which assumes negligible $CP$ violation in $B_s^0$ mixing.

\section{First $B_s^0\to D_{s1}^-(2536) \mu^+ \nu X$ Observation}

A significant fraction of $B_s^0$ mesons decay semileptonically to
orbitally excited $P$-wave $D_s^{**}$ mesons.  Measurements of such
exclusive semileptonic branching fractions are useful for comparisons
of inclusive and exclusive decay rates, the extraction of CKM matrix
elements, $B_s^0$ mixing analyses, and studies of theoretical hadronic
form-factor models.  The \DZero\ collaboration recently made the first
observation of the decay \BsDs1\ in 1.3~fb$^{-1}$ of $p\bar{p}$
collisions at $\sqrt{s} = 1.96~$TeV~\cite{DZero:BsDs1}.

\begin{figure}[h]
\centering
\includegraphics[width=80mm]{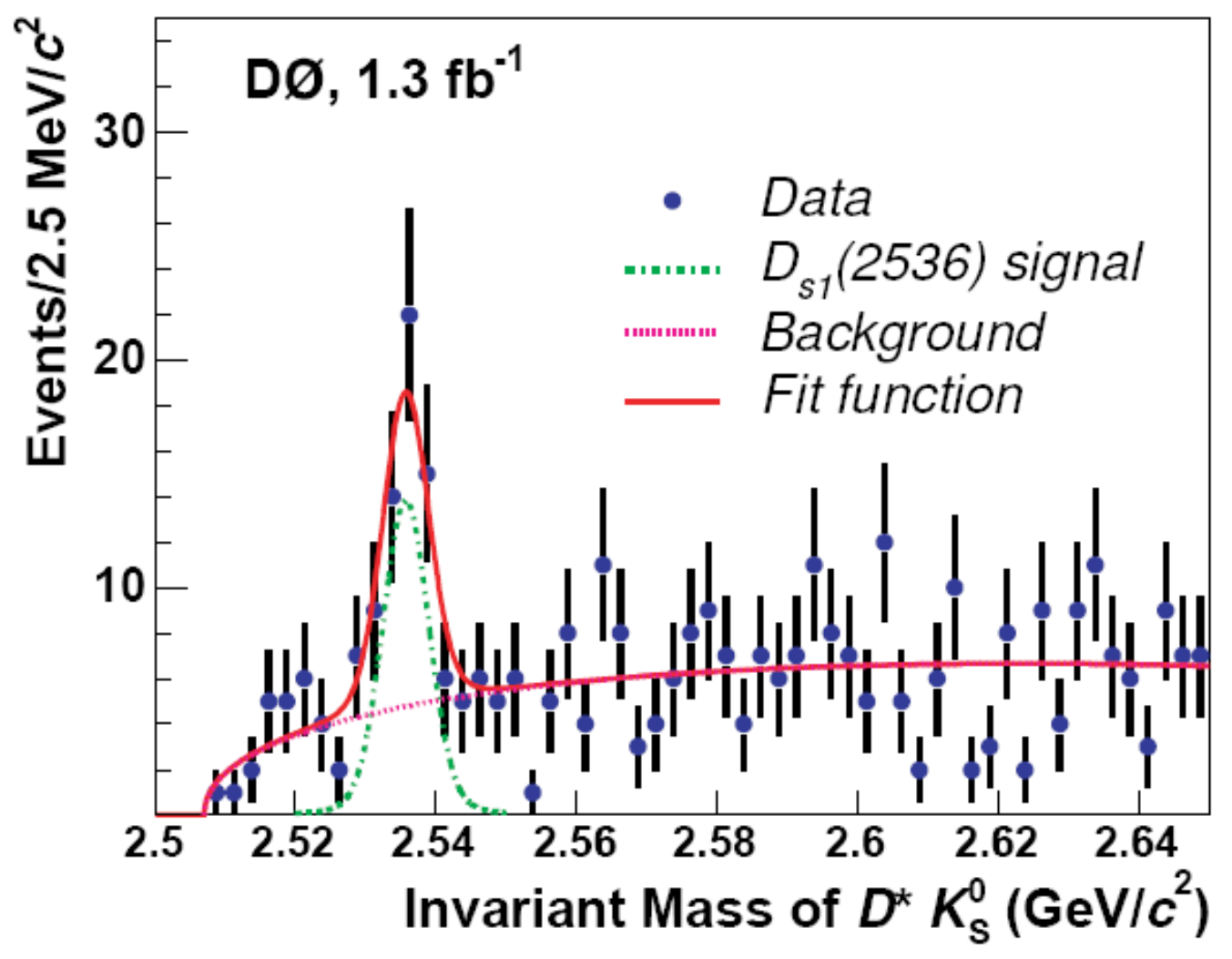}
\caption{Invariant mass distribution of $D^*\, K_S$ candidates with an
  associated muon candidate~\cite{DZero:BsDs1}.  The result of the fit
  to the signal and background is shown by the
  curves.}\label{fig:DZero_BsDs1}
\end{figure}

Based on a sizeable sample of $87\, 506\pm 496 ({\rm stat.})$ $D^*$
meson candidates in an inclusive muon dataset, candidate $D^*\, K_S$
invariant mass combinations were constructed as illustrated in
Fig.~\ref{fig:DZero_BsDs1}.  The indicated fit determined a
$D_{s1}(2536)$ yield of $45.9\pm 9.1 ({\rm stat.})$ candidates with a
statistical significance of 6.1 standard deviations, and a mass of the
$D_{s1}(2536)$ candidate of $2535.7\pm 0.6 ({\rm stat.}) \pm 0.5 ({\rm
  syst.})$~MeV/$c^2$, which was consistent with the PDG world-average
value~\cite{PDG}.  Assuming that $D^-_{s1}(2536)$ production in
semileptonic decays arises solely from $B^0_s$ meson decays, the
\DZero\ collaboration measured the following product of fragmentation
and branching fractions: $f_s\cdot {\cal
  B}($\BsDs1$)\cdot$\calB$D^-_{s1}(2536)\to D^{*-}\, K_S) = \left[
  2.66\pm 0.52 ({\rm stat.}) \pm 0.45 ({\rm syst.})  \right]\times
10^{-4}$.  The derivation of an absolute \calB\BsDs1$)$ branching
fraction from this measured product is useful for comparisons with
theoretical calculations, but requires an assumption for the
$D_{s1}(2536)$ daughter branching fraction and is degraded by the
sizeable world-average experimental uncertainty on $f_s$~\cite{PDG}.

\section{Recent $B_s^0$ Radiative Decay Results}

One-loop effective flavour-changing neutral-current radiative penguin
decays are well known as possible venues for new physics beyond the
Standard Model.  Hitherto unobserved particles can first reveal their
existence in the virtual loop propagator by modifying the Standard
Model couplings, and hence decay rates, in measurable ways.  In the
following, studies of two radiative penguin $B_s^0$ decay modes
conducted by the Belle collaboration using 23.6~fb$^{-1}$ of
asymmetric $e^+e^-$ $\Upsilon(5S)$ collision
data~\cite{Belle:Bsgamma} are reviewed.

\subsection{First \Bsphig\ Observation}
\label{Bsphig}
\begin{figure}[h]
\centering
\includegraphics[width=80mm]{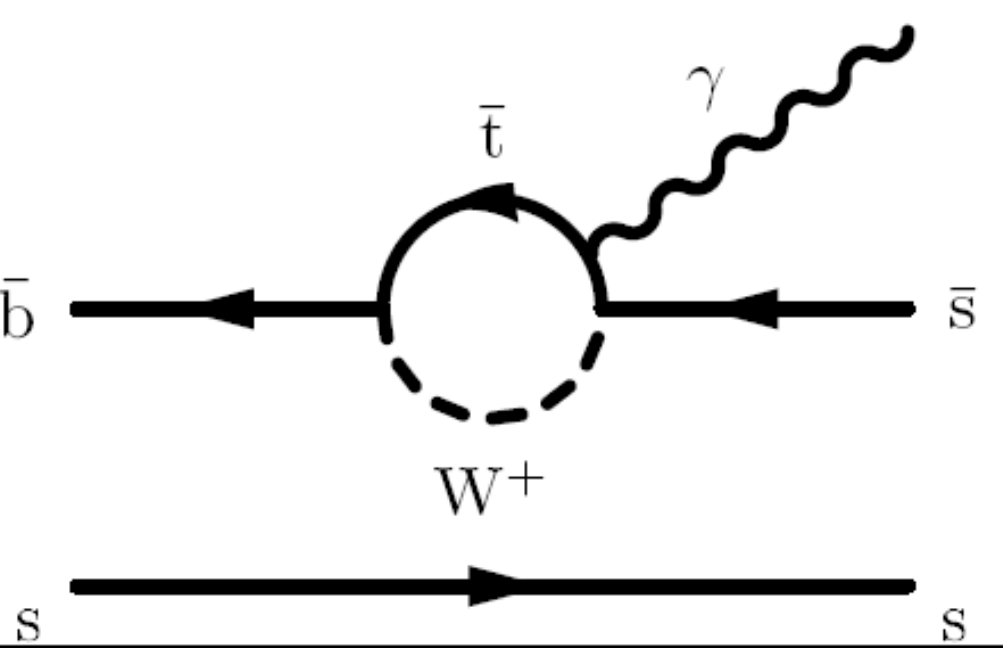}
\caption{Diagrammatic representation of the most probable decay
  process for the
  \Bsphig\ mode~\cite{Belle:Bsgamma}.}\label{fig:Bsphig_diagram}
\end{figure}

The decay \Bsphig\ is described by the Standard Model using a one-loop
radiative penguin diagram, as indicated in
Fig.~\ref{fig:Bsphig_diagram}.  The mode may be considered to be the
strange analogue of the $B\to K^*(892)\,\gamma$ decays, which provided
the first explicit observation of penguin
processes~\cite{CLEO:penguin}.  The subsequent observed agreement
between numerous experimental results for $b\to s\,\gamma$ rates and
Standard Model expectations provides a strong theoretical constraint
also for the analogous \Bsphig\ decay discussed here.

\begin{figure}[h]
\centering
\includegraphics[width=80mm]{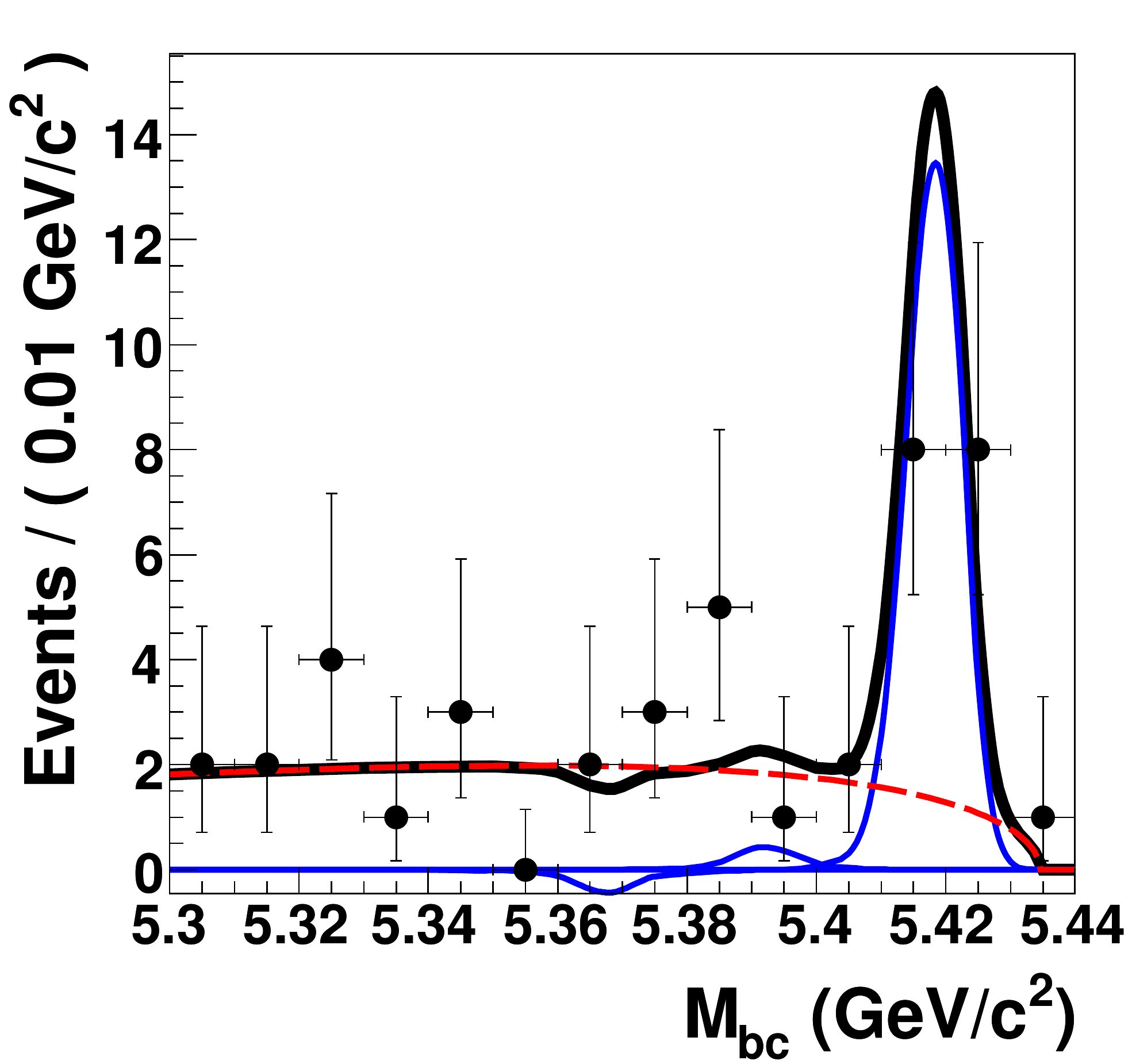}
\caption{The beam-energy-constrained mass $M_{\rm bc}$ projection and
  fit results~\cite{Belle:Bsgamma}.  The points represent data, the
  thick (black) solid curve is the fit function, the thin (blue) curve
  is the signal function, and the dashed (red) curve represents the
  continuum contribution.}\label{fig:Belle_phig_mbc}
\end{figure}

\begin{figure}[h]
\centering
\includegraphics[width=80mm]{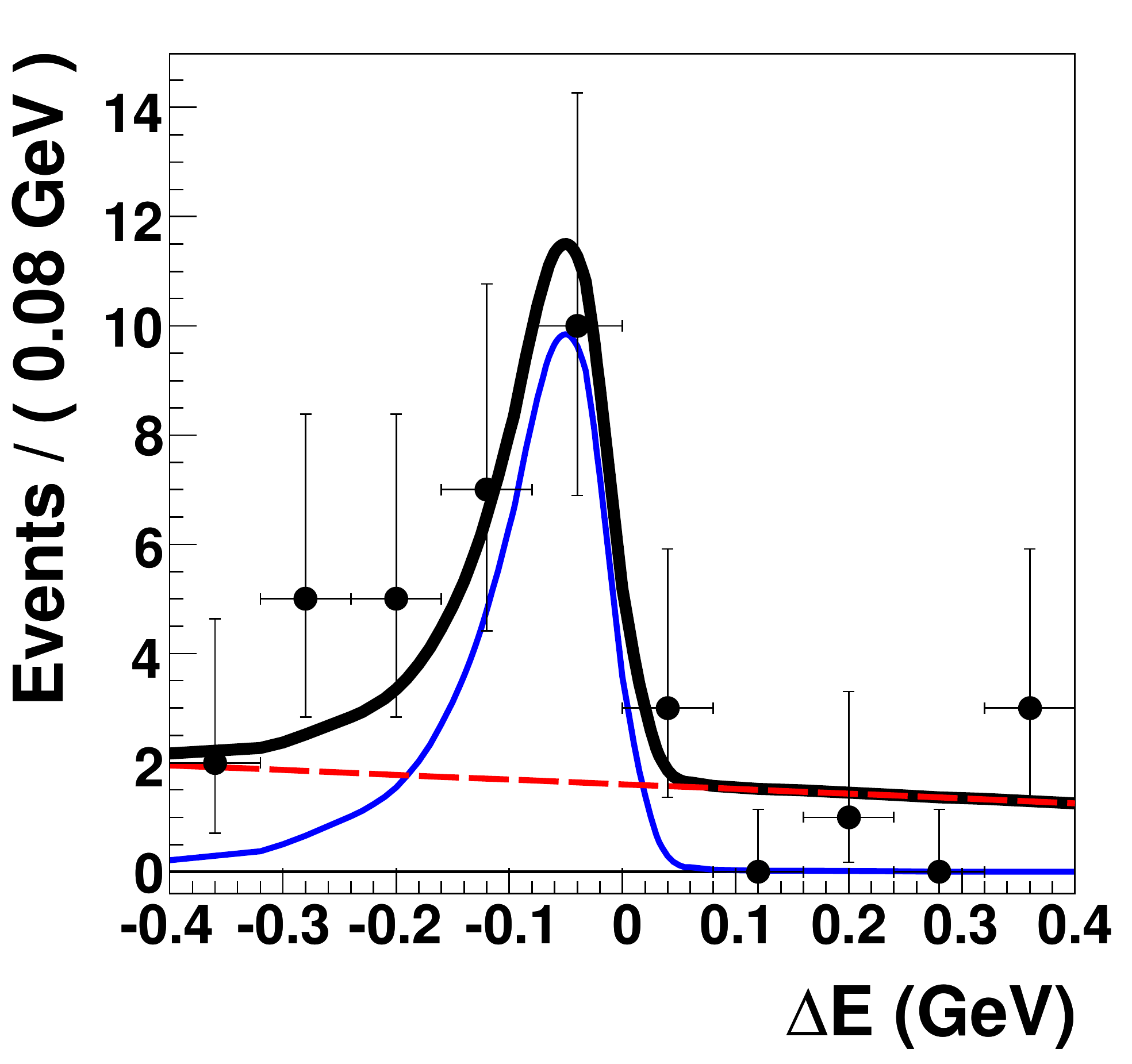}
\caption{The energy difference $\Delta E$ projection and fit
  results~\cite{Belle:Bsgamma}.  The curves and markers are as
  described in
  Fig.~\ref{fig:Belle_phig_mbc}.}\label{fig:Belle_phig_dE}
\end{figure}

\begin{figure}[h]
\centering
\includegraphics[width=80mm]{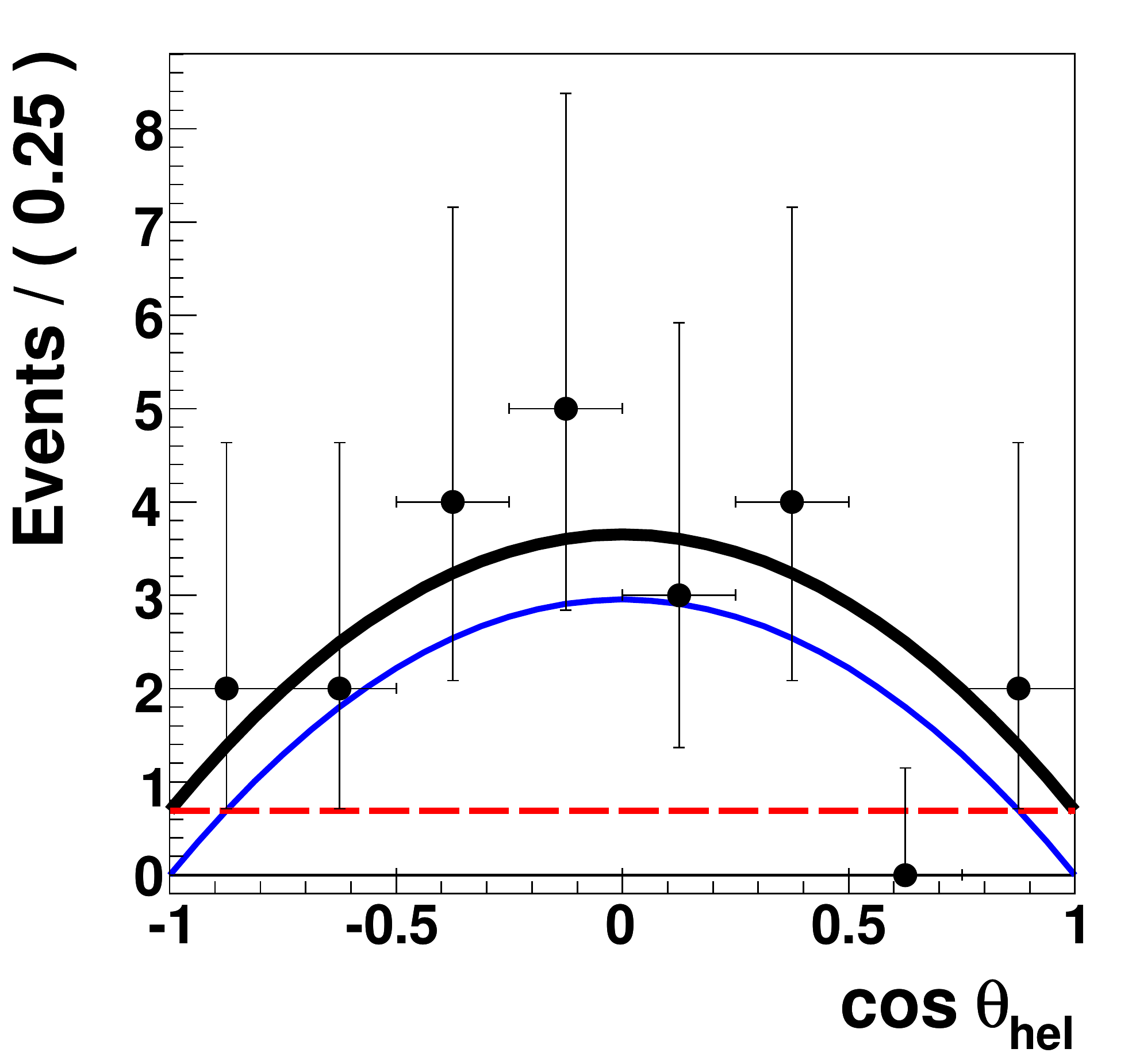}
\caption{The $\cos\theta_{\rm hel}$ projection and fit
  results~\cite{Belle:Bsgamma}.  The curves and markers are as
  described in
  Fig.~\ref{fig:Belle_phig_mbc}.}\label{fig:Belle_phig_costh}
\end{figure}

The Belle analysis~\cite{Belle:Bsgamma} made use of a
three-dimensional unbinned maximum likelihood fit involving
observables of beam-energy-constrained mass $M_{\rm bc}$, energy
difference $\Delta E$, and the cosine of the helicity angle
$\cos\theta_{\rm hel}$, where $\theta_{\rm hel}$ was the angle between
the $B_s^0$ and $K^+$ mesons in the $\phi$ meson rest frame.
Projections on to these three observables and the fit results are
illustrated in Figs.~\ref{fig:Belle_phig_mbc},
\ref{fig:Belle_phig_dE}, and \ref{fig:Belle_phig_costh}.

The fit resulted in $18^{+6}_{-5}$ signal candidates with a
statistical significance of 5.5 standard deviations, and a first
observed branching fraction of \calB\Bsphig$) = \left[
  57\,^{+18}_{-15} ({\rm stat.}) \, ^{+12}_{-11} ({\rm syst.})\right]
\times 10^{-6}$ was determined~\cite{Belle:Bsgamma}.

\subsection{Search for Radiative \Bsgg\ Decay}

\begin{figure}[h]
\centering
\includegraphics[width=80mm]{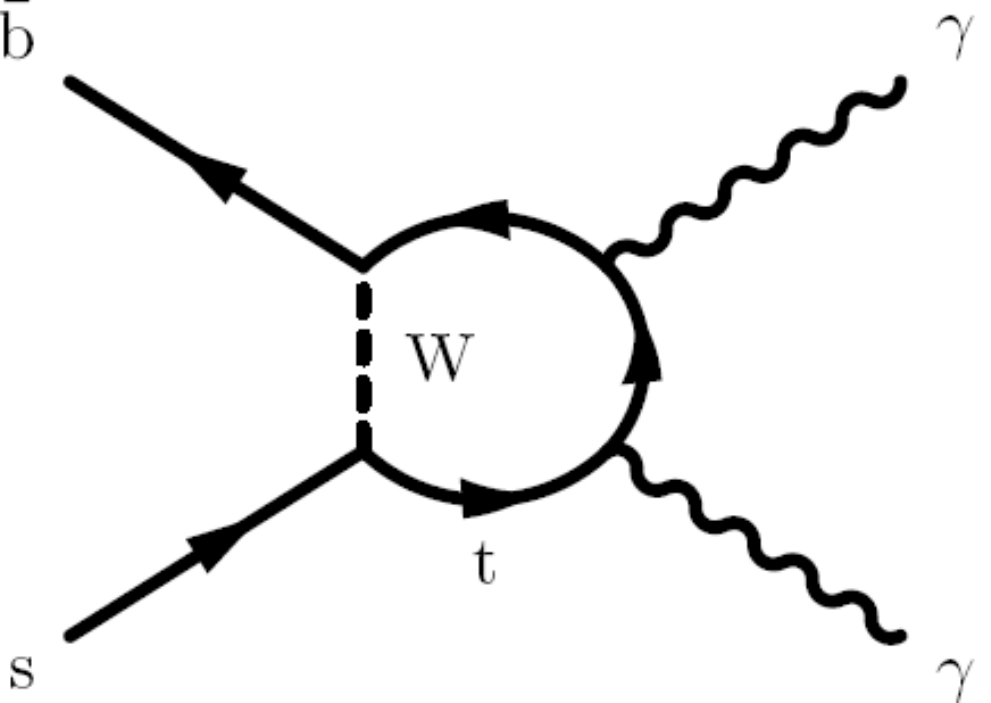}
\caption{Diagrammatic representation of the most probable decay
  process for the \Bsgg\ mode~\cite{Belle:Bsgamma}
  .}\label{fig:Bsgg_diagram}
\end{figure}

As for the \Bsphig\ mode, the \Bsgg\ channel, a diagram for which is
depicted in Fig.~\ref{fig:Bsgg_diagram}, is similarly expected to be
largely constrained by the known $B\to X_s\,\gamma$ branching
fraction~\cite{Bertolini}; however, various new physics sources have
been suggested that could enhance the \Bsgg\ branching fraction by up
to an order of magnitude without compromising this constraint.  These
include a two Higgs doublet model with flavour-changing neutral
currents~\cite{Aliev}, a fourth-quark generation~\cite{Huo}, and
supersymmetry with broken $R$-parity~\cite{Gemintern}.

\begin{figure}[h]
\centering
\includegraphics[width=80mm]{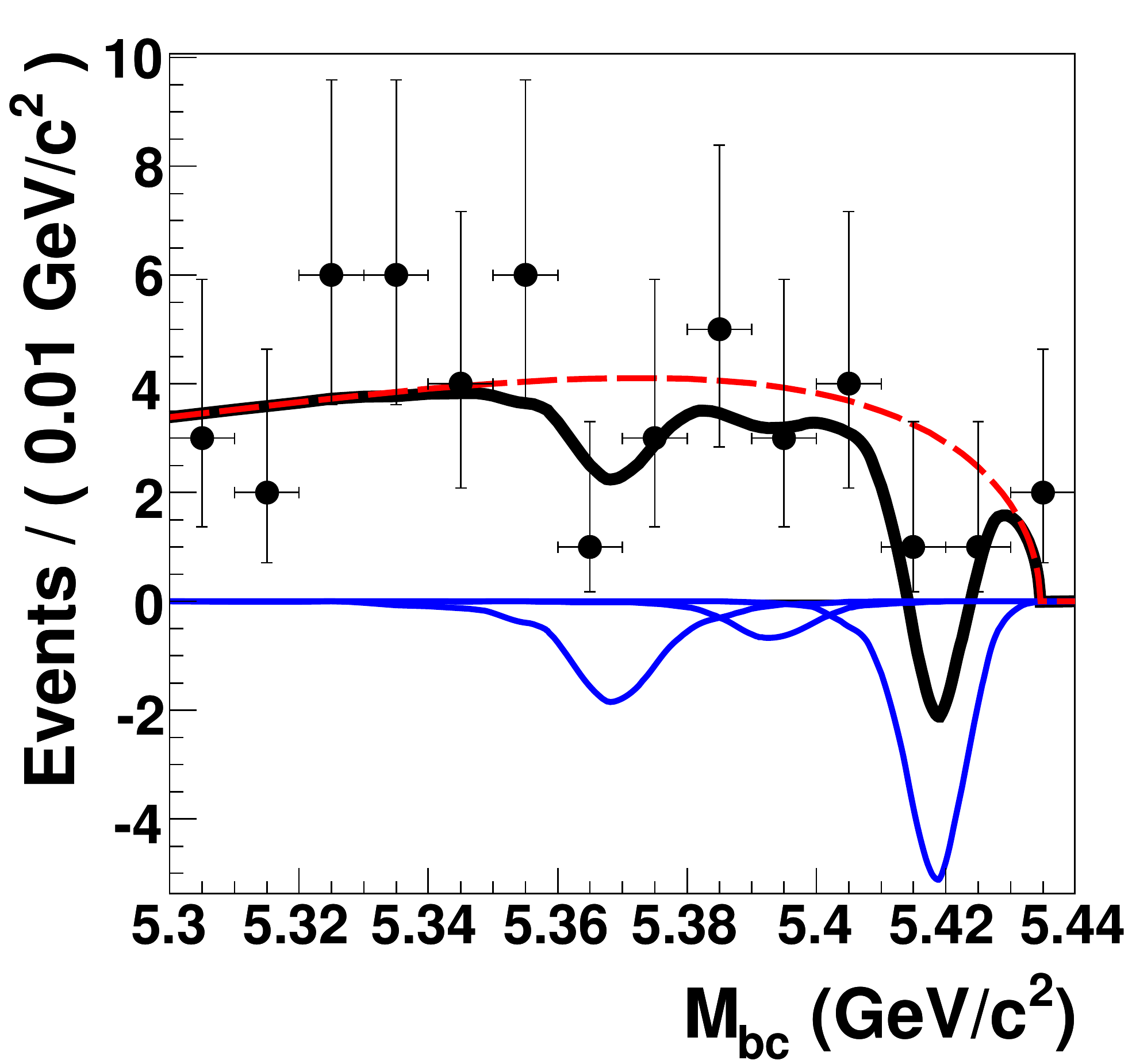}
\caption{The beam-energy-constrained mass $M_{\rm bc}$ projection and
  fit results~\cite{Belle:Bsgamma}.  The points represent data, the
  thick (black) solid curve is the fit function, the thin (blue) curve
  is the signal function, and the dashed (red) curve represents the
  continuum contribution.}\label{fig:Belle_gg_mbc}
\end{figure}

\begin{figure}[h]
\centering
\includegraphics[width=80mm]{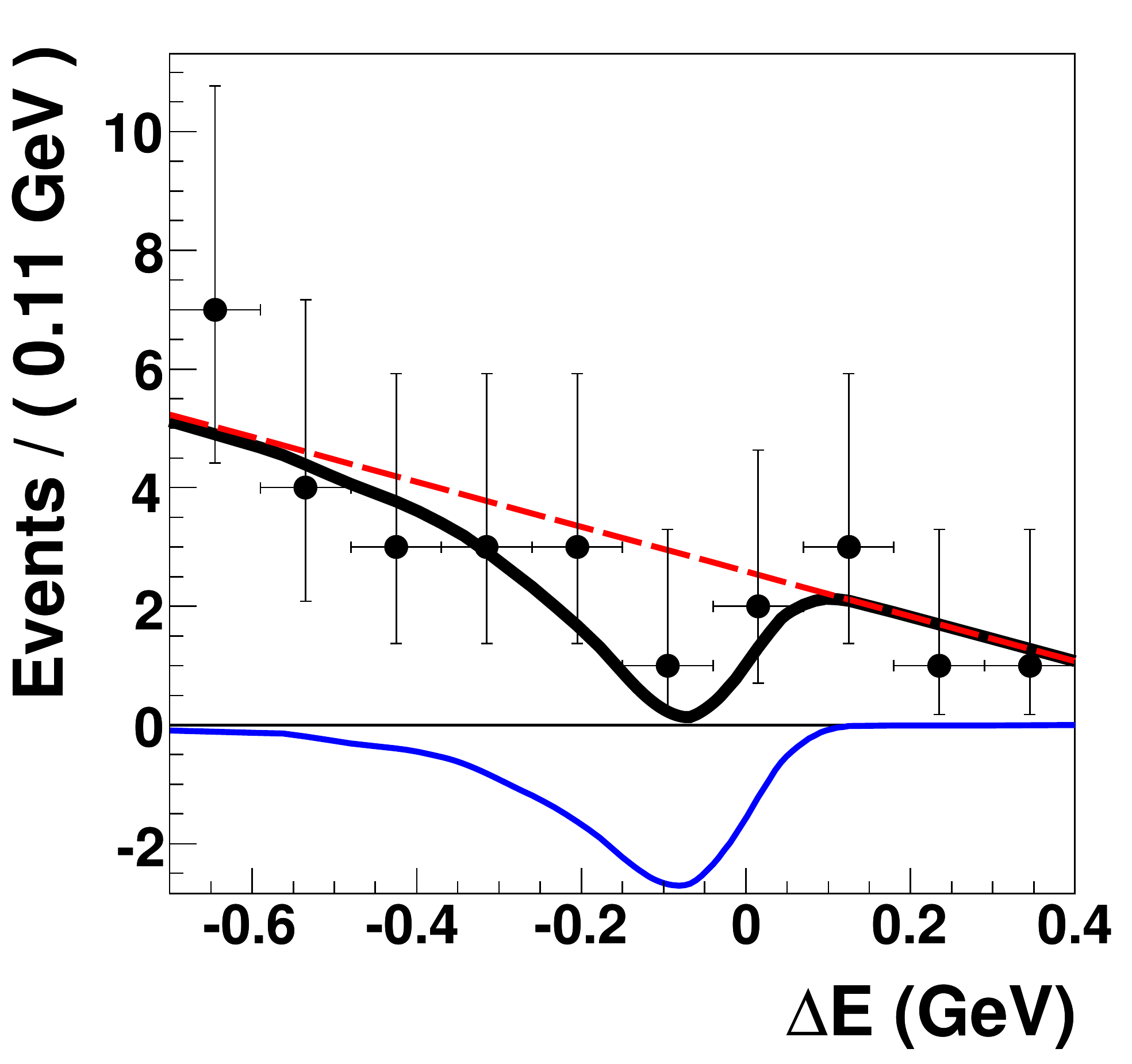}
\caption{The energy difference $\Delta E$ projection and fit
  results~\cite{Belle:Bsgamma}.  The curves and markers are as
  described in Fig.~\ref{fig:Belle_gg_mbc}.}\label{fig:Belle_gg_dE}
\end{figure}

The Belle \Bsgg\ analysis consisted of a two-dimensional unbinned
maximum likelihood fit involving the $M_{\rm bc}$ and $\Delta E$
observables previously defined in Sect.~\ref{Bsphig}.
Figs.~\ref{fig:Belle_gg_mbc} and \ref{fig:Belle_gg_dE} depict the
projections on to the two observables as well as the fit results.

There being no significant \Bsgg\ signal observed, the Belle
collaboration set a 90\% confidence level branching-fraction limit of
\calB\Bsgg$) < 8.7 \times 10^{-6}$, which, though six times more
restrictive than the previous limit~\cite{Belle:BsDssDss}, still
stands at least an order of magnitude higher than the Standard Model
and new physics predictions.

\section{Charmless Two-Body $B_s^0$ Modes}
\label{Bshh}
Two-body charmless hadronic decays of neutral $b$ hadrons can provide
insight into both the CKM matrix and possible new physics phenomena.
The CDF experiment can reconstruct significant samples of
these decay modes by virtue of the high yields of $b$-quark production
at the Fermilab Tevatron $p\bar{p}$ collider and the use of a
dedicated trigger on impact parameters of charged-particle tracks,
discussed in Sect.~\ref{BsDsK}.

The degree of penguin-tree interference at play in these decay modes
depends on the CKM angle $\gamma$ ($\phi_3$), hadronic amplitudes and
phases, and possibly the presence of new physics.  Strategies for
understanding this fertile system of processes involve the study of
several modes related by isospin and $SU(3)$ flavour and the
combination of multiple observables such as branching fractions and
$CP$ asymmetries.

The CDF collaboration recently updated its suite of $B^0_{(s)}\to
h^+\, h^{\prime -}$ branching-fraction results based on 1~fb$^{-1}$ of
$p\bar{p}$ collisions at $\sqrt{s} = 1.96~$TeV~\cite{CDF:Bshh}, where
$h$ denotes either a $\pi$ or $K$ meson.  The main change, which
consisted of a better determination of the relative isolation
efficiencies for $B^0$ and $B_s^0$ candidates, resulted in improved
branching-fraction measurements for the $B_s^0$ modes.  The results,
for which the largest systematic uncertainties arose due to $dE/dx$
particle-identification aspects, are summarized in
Table~\ref{tab:Bshh} and compared with a set of theoretical
predictions~\cite{Beneke}, which shows good agreement.  Another more
recent set of predictions due to Chiang and Zhou~\cite{Chiang} also
shows reasonable agreement.  More theoretical calculations are cited
in Ref.~\cite{CDF:Bshh_180pb}.

\begin{table*}[t]
\begin{center}
\caption{Updated $B_s^0\to h^+\, h^{\prime -}$ signal yields, ratios
  of product fragmentation and branching-fraction measurement results,
  and derived absolute branching-fraction results from
  CDF~\cite{CDF:Bshh} with input from the HFAG~\cite{HFAG}.  Where
  applicable, the first uncertainties are statistical and the second
  are systematic.  The rightmost column represents an example
  theoretical prediction by Beneke and Neubert~\cite{Beneke} for
  comparison with the measured results.  The ranges shown for the
  theoretical prediction represent coverage of multiple theory
  parameter fit scenarios; theoretical uncertainties are not
  indicated.}
\begin{tabular}{|l|l|l|c|c|}
\hline \multicolumn{1}{|c|}{\textbf{Decay}} &
\multicolumn{1}{c|}{\textbf{Signal}} &
\multicolumn{1}{c|}{\textbf{CDF}} & \multicolumn{2}{c|}{\textbf{${\cal
      B}$ $\times \, 10^6$}} \\
\multicolumn{1}{|c|}{\textbf{Mode}} &
\multicolumn{1}{c|}{\textbf{Yield}} &
\multicolumn{1}{c|}{\textbf{Measurement~\cite{CDF:Bshh}}} &
\textbf{CDF Derived} & \textbf{Beneke \& Neubert~\cite{Beneke}}
\\ [3pt]
\cline{1-5} & & & & \\

$B^0_s\to K^+\, K^-$ & $1307\pm 64$ & $\frac{f_s}{f_d}\cdot\frac{{\cal
    B}(B^0_s\to K^+\, K^-)}{{\cal B}(B^0\to K^+\, \pi^-)} = 0.347\pm
0.020\pm 0.021$ & $25.8\pm 1.5\pm 3.9$ & $28 \to 36$ \\ [5pt]

$B^0_s\to K^-\, \pi^+$ & $230\pm 34\pm 16$ & $\frac{f_s}{f_d}\cdot\frac{{\cal
    B}(B^0_s\to K^-\, \pi^+)}{{\cal B}(B^0\to K^+\, \pi^-)} = 0.071\pm
0.010\pm 0.007$ & $5.27\pm 0.74\pm 0.90$ & $6.8 \to 10.4$ \\ [5pt]

$B^0_s\to \pi^+\, \pi^-$ & $26\pm 16\pm 14$ & $\frac{f_s}{f_d}\cdot\frac{{\cal
    B}(B^0_s\to \pi^+\, \pi^-)}{{\cal B}(B^0\to K^+\, \pi^-)} = 0.007\pm
0.004\pm 0.005$ & $< 1.3$ @ 90\% CL & $0.027 \to 0.155$ \\ [5pt]
\hline
\end{tabular}
\label{tab:Bshh}
\end{center}
\end{table*}

For meaningful comparisons of experiment and theory, however, it is
important that the scope be widened to include also the $B^+$ and
$B^0$ branching fractions, as well as the $CP$ asymmetries.
Inconsistencies between the current landscape of experimental results
and theoretical predictions suggest that there may be more $SU(3)$
breaking than expected in the strong phases or that there has been a
breakdown of QCD factorization~\cite{Chiang_Gronau_Rosner}.

\section{First $\Lambda_b^0\to p\, (\pi^-,K^-)$ Observations}

No $CP$-violating decay-rate asymmetries have previously been measured
in baryon decays.  Two-body charmless decays of $\Lambda_b^0$ baryons
with a final-state proton and a charged pion or kaon are self tagging
and therefore promising vehicles for the study of $CP$-violating
asymmetries~\cite{Mohanta_Giri_Khanna}.  CDF is currently the only
experiment able to collect and reconstruct such two-body $b$-baryon
decay candidates~\cite{CDF:Lambdab_pKpi}.

Branching fractions of \Lbppi\ and \Lbpk\ decays have also been shown
to have possible sensitivity to new physics.  Mohanta has provided a
prediction suggesting that Minimal Supersymmetric Extensions of the
Standard Model in which $R$-parity is violated could enhance these
branching fractions by a factor of ${\cal O}(100)$~\cite{Mohanta}.

\begin{figure}[h]
\centering
\includegraphics[width=80mm]{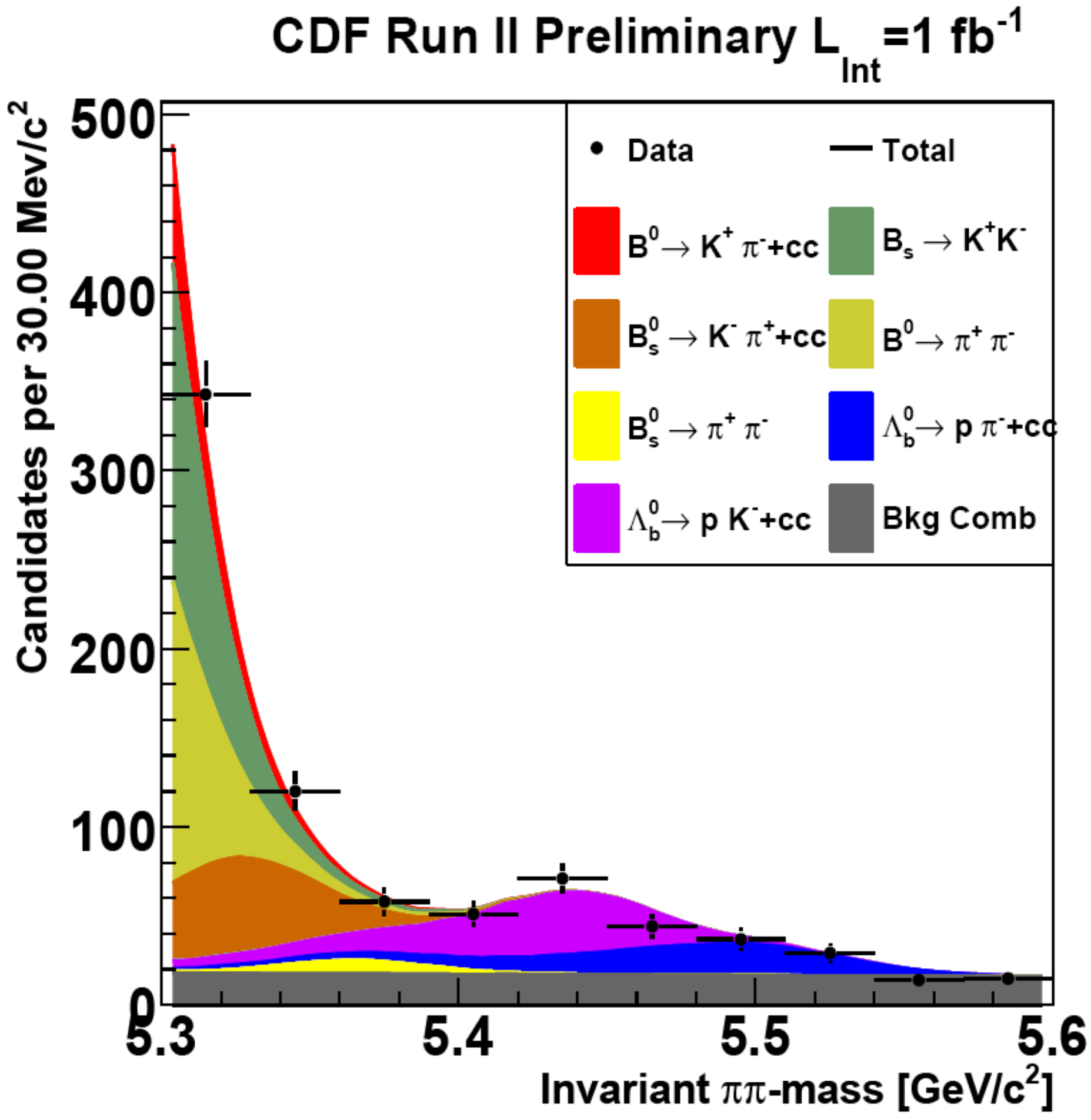}
\caption{The dipion invariant mass $M_{\pi\pi}$ projection and fit
  results~\cite{CDF:Lambdab_pKpi}.  The signal \Lbppi\ and
  \Lbpk\ contributions are indicated in blue and magenta,
  respectively.}\label{fig:Lambda_b_pKpi}
\end{figure}

The analysis approach employs the same trigger and fit machinery as
that used to obtain the results reported in Sect.~\ref{Bshh} above,
but has been optimized specifically for dedicated measurements of
\Lbppi\ and \Lbpk\ branching fractions (and $CP$ asymmetries, though
these lie beyond the scope of this review).  Details of the candidate
selection criteria are provided in Ref.~\cite{CDF:Lambdab_pKpi}.  A
fit is performed in terms of a particle-identification observable and
three kinematic variables: an invariant mass calculated with a pion
hypothesis assigned to both charged tracks; a signed momentum
imbalance $\alpha \equiv (1 - p_1/p_2)q_1$, where $p_1$ ($p_2$) is the
lower (higher) of the particle momenta and $q_1$ is the sign of the
charge of the track with momentum $p_1$; and the scalar sum of the two
track momenta.  A projection of the dipion invariant mass observable
and the fit results is illustrated in Fig.~\ref{fig:Lambda_b_pKpi}.

The signal yields were corrected for efficiencies and related to the
abundant $B^0\to K^+\,\pi^-$ mode, whose branching fraction is known
with $B$-factory precision~\cite{PDG}.  The preliminary observed
branching-fraction ratios, measured in a product with the ratios of
production cross sections and fragmentation fractions,
were~\cite{CDF:Lambdab_pKpi}

\begin{eqnarray}
\frac{\sigma(p\bar{p}\to\Lambda_b^0\, X, p_T > 6\, {\rm
    GeV/}c)}{\sigma(p\bar{p}\to B^0\, X, p_T > 6\, {\rm GeV/}c)}
\frac{{\cal B}(\Lambda_b^0\to p\,\pi^-)}{{\cal B}(B^0\to K^+\,\pi^-)}
 \nonumber \\
= \,\, 0.0415\pm 0.0074 ({\rm stat.})\pm 0.0058 ({\rm syst.}) \nonumber
\end{eqnarray}
and
\begin{eqnarray}
\frac{\sigma(p\bar{p}\to\Lambda_b^0\, X, p_T > 6\, {\rm
    GeV/}c)}{\sigma(p\bar{p}\to B^0\, X, p_T > 6\, {\rm GeV/}c)}
\frac{{\cal B}(\Lambda_b^0\to p\,K^-)}{{\cal B}(B^0\to K^+\,\pi^-)}
 \nonumber \\
= \,\, 0.0663\pm 0.0089 ({\rm stat.})\pm 0.0084 ({\rm syst.}), \nonumber
\end{eqnarray}
for which the candidate isolation efficiencies constituted the dominant
source of systematic uncertainty.  Preliminary measurements of the
$CP$ decay-rate asymmetries for the \Lbppi\ and \Lbpk\ modes were also
computed and are reported in Ref.~\cite{CDF:Lambdab_pKpi}.

World-average values of the poorly known ratio of fragmentation
fractions $f_{b{\rm -baryon}} / f_d$ and the $B^0\to K^+\,\pi^-$ decay
rate~\cite{PDG} were used to extract the derived absolute branching
fractions~\cite{CDF:Lambdab_pKpi}

\begin{eqnarray}
{\cal B}(\Lambda_b^0\to p\,\pi^-)
= \left[3.1\pm 0.6 ({\rm stat.}) \pm  0.7 ({\rm syst.}) \right]\times
10^{-6} \nonumber
\end{eqnarray}
and 
\begin{eqnarray}
{\cal B}(\Lambda_b^0\to p\, K^-)
= \left[5.0\pm 0.7 ({\rm stat.}) \pm  1.0 ({\rm syst.}) \right]\times
10^{-6}, \nonumber
\end{eqnarray}
which agree with Standard Model expectations and exclude the ${\cal
  O}(10^{-4})$ branching fractions predicted by the $R$-parity
violating Minimal Supersymmetric extension of the Standard
Model~\cite{Mohanta}.

\section{Conclusions}

Recent studies of $p\bar{p}$ and $e^+e^-$ collisions by the Belle,
CDF, and \DZero\ collaborations have yielded a plethora of new decay
observations and branching-fraction measurements for $B_s^0$ mesons
and $\Lambda_b^0$ baryons.  First observations and new results in
three classes of $B_s^0 \to D_s^{(*(*))}\, X$ modes have recently been
reported; the first radiative penguin in the $B_s^0$ system, \Bsphig,
has been observed and the branching fraction measured; a significantly
reduced upper limit has been determined for the radiative \Bsgg\ mode,
which has been deemed sensitive to a number of possible new physics
scenarios; $B_s^0\to h^-\, h^{\prime +}$ decay rates, which are
crucial to the formation of a consistent picture between experiment
and theory in the larger space of two-body charmless $b$-hadron decay
rates and $CP$ asymmetries, have been updated; and first observations
of the modes \Lbppi\ and \Lbpk\ have been made and their branching
fractions preliminarily measured.

A recurring aspect in the study of $B_s^0$ and $b$-baryon decay rates
at $p\bar{p}$ and $e^+e^-$ colliders is the reliance on $f_s(b\to
B^0_s)$ and $f_{b{\rm -baryon}}(b\to\Lambda_b^0)$ fragmentation
fractions for interpretation of the results.  These fragmentation
fractions, in particular their dependence on the $b$-quark momentum
and the accelerator collision environment, are poorly understood and
have so far been measured with sizeable uncertainties~\cite{PDG}.
Improved measurements of $f_s$ and $f_{b{\rm -baryon}}$ are therefore
crucial to this important heavy $b$-hadron frontier of flavour
physics.

Given the recent results, further study with additional data is
strongly warranted to hunt for signs of new physics lying beyond the
Standard Model, to seek answers to theoretical questions about
Standard-Model manifestations of electroweak and QCD phenomenology, to
employ flavour-tagging techniques, and to continue exploration of this
rich and exciting sector of flavour particle physics.  Further to the
results presented in this review, an additional $2-3\times$ factor of
time-integrated luminosity is expected to be analyzed from the
Fermilab Tevatron, the LHC$b$ experiment is expected to begin
datataking soon at CERN, and a high-intensity Super-$B$ facility is
envisaged in the coming years.

\begin{acknowledgments}
Colleagues in the Belle, CDF, and \DZero\ collaborations, as well as
the numerous technical and accelerator personnel at Fermilab and KEK,
are acknowledged for their vital contributions to the results
described in this review.  The FPCP 2008 organizers are also warmly
thanked for arranging an enjoyable and stimulating meeting at the
National Taiwan University in Taipei.
\end{acknowledgments}

\bigskip 

\end{document}